\documentclass{article}
\usepackage{ijcai25}

\usepackage{times}
\usepackage{soul}
\usepackage{url}
\usepackage[hidelinks]{hyperref}
\usepackage[utf8]{inputenc}
\usepackage[small]{caption}
\usepackage{graphicx}
\usepackage{amsmath}
\usepackage{amsthm}
\usepackage{booktabs}
\usepackage{algorithm}
\usepackage{algorithmic}
\usepackage[switch]{lineno}

\usepackage{subcaption}
\usepackage[capitalise,noabbrev]{cleveref}
\usepackage{svg}

\title{Will Systems of LLM Agents Cooperate: An Investigation into a Social Dilemma}

\author{
Richard Willis$^1$
\and
Yali Du$^1$\and
Joel Z Leibo$^{1,2}$\And
Michael Luck$^3$\\
\affiliations
$^1$King's College London\\
$^2$Google DeepMind\\
$^3$University of Sussex\\
\emails
richard.willis@kcl.ac.uk,
yali.du@kcl.ac.uk,
jzl@deepmind.com,
michael.luck@sussex.ac.uk
}

\begin{document}

\maketitle

\begin{abstract}
As autonomous agents become more prevalent, understanding their collective behaviour in strategic interactions is crucial. This study investigates the emergent cooperative tendencies of systems of Large Language Model (LLM) agents in a social dilemma. Unlike previous research where LLMs output individual actions, we prompt state-of-the-art LLMs to generate complete strategies for iterated Prisoner's Dilemma. Using evolutionary game theory, we simulate populations of agents with different strategic dispositions (aggressive, cooperative, or neutral) and observe their evolutionary dynamics. Our findings reveal that different LLMs exhibit distinct biases affecting the relative success of aggressive versus cooperative strategies. This research provides insights into the potential long-term behaviour of systems of deployed LLM-based autonomous agents and highlights the importance of carefully considering the strategic environments in which they operate.
\end{abstract}

\section{Introduction}
The increasing deployment of autonomous agents based on Large Language Models (LLMs) \cite{wang24__a_survey_on_large_language_model_based_autonomous_agents} in real-world applications necessitates an examination of their collective impact on machine-machine interactions and human culture \cite{brinkmann23__machine_culture}.
Whilst individual LLM capabilities are frequently assessed, understanding their collective behaviours and societal consequences remains crucial and underexplored.

The development of social capabilities in these agents may lead to dual-use skills usable for both pro-social and anti-social purposes, termed \emph{differential capabilities}. \cite{dafoe20__open_problems_in_cooperative_ai}.
This duality raises questions about the balance between cooperation and conflict in autonomous agent interactions.
Furthermore, situations such as social dilemmas pose inherent risks, as competent agents acting rationally can lead to suboptimal collective outcomes \cite{pan23__do_the_rewards_justify_the_means_measuring_tradeoffs_between_rewards_and_ethical_behavior_in_the_machiavelli_benchmark}.
If agents succeed through aggressive behaviours, competitive pressures could potentially drive systems towards suboptimal equilibria \cite{anwar24__foundational_challenges_in_assuring_alignment_and_safety_of_large_language_models}.

Prior assessments of LLMs have evaluated their capacity to engage in various multiplayer games \cite{mao23__alympics,yocum23__mitigating_generative_agent_social_dilemmas,park23__generative_agents,gong23__mindagent,zhang24__proagent,wu23__chatarena} and the emergent behaviours of systems of LLM agents has been explored.
Conventionally, however, LLMs are prompted to output a single action in response to a given game state or trajectory.
Recent analyses have revealed that LLMs struggle when tasked with making decisions at this level of granularity \cite{fan24__can_large_language_models_serve_as_rational_players_in_game_theory}.
In such scenarios, they fail to identify basic patterns, such as an opponent mirroring their own moves.
This limitation likely stems from the fact that LLMs are not specifically trained for data science tasks, or to handle inputs of this format.

In response, in contrast to prior work, our we prompt LLMs to create fixed strategies in natural language, which are subsequently implemented as algorithms in Python.
This method enables the LLMs to craft their approach at a higher level of abstraction.
For example, with our approach, we observe that many LLM strategies utilise pattern recognition and successfully implement code to detect simple patterns up to a fixed length.
A key advantage of creating strategies to encode as algorithms, rather than outputting individual actions, is that it facilitates behaviour checking in advance.
This approach allows users to inspect the strategy, test for safety and robustness, and explore the potential implications prior to deployment.

Our research employs the iterated Prisoner's Dilemma (IPD) \cite{axelrod80__effective_choice_in_the_prisoners_dilemma,crandall14__towards_minimizing_disappointment_in_repeated_games,beaufils96__our_meeting_with_gradual} to evaluate the balance between pro-social and anti-social behaviours exhibited by state-of-the-art LLM agents.
We utilise evolutionary game theory \cite{axelrod86__an_evolutionary_approach_to_norms,mahmoud10__an_analysis_of_norm_emergence_in_axelrods_model,nowak04__emergence_of_cooperation_and_evolutionary_stability_in_finite_populations,nowak06__five_rules_for_the_evolution_of_cooperation} to investigate whether systems of frontier LLM agents are predisposed to exhibit cooperation or conflict under competitive pressures.
The choice of IPD provides a robust mathematical framework for analysing the strategic behaviour and cooperative biases of LLM agents.
Moreover, as game theory represents a high-level abstraction of various social phenomena, with applications spanning economics, politics, sociology, and psychology, insights gained from LLM performance in these scenarios may have far-reaching implications across multiple disciplines.

Our contributions are as follows: we quantify the relative success of pro-social and anti-social LLM agent behaviours in a social dilemma; we assess the relative likelihoods of systems of LLM agents converging to anti-social or pro-social equilibria; and, we release our code\footnote{\url{https://github.com/willis-richard/evollm}} as an evaluation suite for model developers to assess the emergent behaviour of their products.

In addition to the above, we provide supplementary analysis: we verify that the LLM strategies exhibit the requested behaviours; we benchmark the LLM strategy performance against human written strategies using the same setup as prior work \cite{beaufils96__our_meeting_with_gradual}; and, we investigate the impact of noisy actions, which represent execution mistakes.

\section{Related Work}
Evaluating and benchmarking the capabilities of LLMs is common practice, as these models frequently exhibit emergent capabilities, such as theory of mind \cite{vanduijn23__theory_of_mind_in_large_language_models} and reasoning \cite{kojima22__large_language_models_are_zeroshot_reasoners}, despite certain limitations in their abilities \cite{sclar23__minding_language_models_lack_of_theory_of_mind,dziri23__faith_and_fate}.
Moreover, LLMs are increasingly utilized in multiagent systems (MAS).
Notably, a line of research has focused on modelling societies of generative agents \cite{park23__generative_agents} and examining their performance in social dilemmas \cite{yocum23__mitigating_generative_agent_social_dilemmas}.
However, we perceive a gap in the assessment of the emergent behaviours of systems of LLMs.
Our proposal aims to expand the evaluation and benchmarking of LLMs to encompass an analysis of their emergent collective behaviours.

LLMs have been used to play games from game theory \cite{aher23__using_large_language_models_to_simulate_multiple_humans_and_replicate_human_subject_studies,horton23__large_language_models_as_simulated_economic_agents,wu23__chatarena}, including iterated normal-form games \cite{akata23__playing_repeated_games_with_large_language_models}, extensive-form games \cite{mao23__alympics}, Markov social dilemma games \cite{yocum23__mitigating_generative_agent_social_dilemmas} and team games \cite{zhang24__proagent,gong23__mindagent}.
These games serve as proxies for real-world behaviours and assess the abilities of LLM agents in a range of scenarios.
However, LLMs can struggle to play games at an action-level granularity \cite{fan24__can_large_language_models_serve_as_rational_players_in_game_theory}.
In contrast, our approach involves having the LLMs output strategies in advance.

LLMs have been suggested for use in game theoretic setting \cite{gemp24__states_as_strings_as_strategies} and modelling human societies and social phenomena \cite{park23__generative_agents,vezhnevets23__generative_agentbased_modeling_with_actions_grounded_in_physical_social_or_digital_space_using_concordia,piatti24__cooperate_or_collapse,dezarza23__emergent_cooperation_and_strategy_adaptation_in_multiagent_systems,gao24__large_language_models_empowered_agentbased_modeling_and_simulation}.
While our approach similarly utilises games to evaluate LLM behaviour, our focus diverges from improving LLM performance.
Instead, we aim to critically assess the balance between aggressive and cooperative behaviours exhibited by these models, and to analyse the emergent dynamics of systems comprising multiple LLM agents with varying behavioural tendencies.

IPD has been extensively employed in various fields of study to model and analyse strategic decision-making in repeated interactions \cite{axelrod80__effective_choice_in_the_prisoners_dilemma,crandall14__towards_minimizing_disappointment_in_repeated_games,rapoport15__is_titfor-tat_the_answer_on_the_conclusions_drawn_from_axelrods_tournaments,press12__iterated_prisoners_dilemma_contains_strategies_that_dominate_any_evolutionary_opponent,knight16__an_open_reproducible_framework_for_the_study_of_the_iterated_prisoners_dilemma,kendall07__the_iterated_prisoners_dilemma,nowak93__a_strategy_of_winstay_lose-shift_that_outperforms_tit-for-tat_in_the_prisoners_dilemma_game}.
Furthermore, researchers have used IPD to study the emergence and stability of cooperative behaviours in populations: 
it has helped explain phenomena such as reciprocal altruism and the evolution of cooperation among non-kin individuals \cite{axelrod86__an_evolutionary_approach_to_norms,mahmoud10__an_analysis_of_norm_emergence_in_axelrods_model,nowak04__emergence_of_cooperation_and_evolutionary_stability_in_finite_populations,nowak06__five_rules_for_the_evolution_of_cooperation,stewart13__from_extortion_to_generosity_evolution_in_the_iterated_prisoners_dilemma,hilbe13__evolution_of_extortion_in_iterated_prisoners_dilemma_games,wahl99__the_continuous_prisoners_dilemma}.
The incorporation of noisy actions into IPD models \cite{wu95__how_to_cope_with_noise_in_the_iterated_prisoners_dilemma,wahl99__the_continuous_prisoners_dilemmaa} serves a dual purpose: it simulates the uncertainty of action outcomes and represents the potential for execution errors by agents.
This added complexity allows us to assess the robustness and adaptability of LLM agent behaviours under more realistic, imperfect conditions.

\section{Method}
\label{sec:method}
We investigate whether LLM agents are more successful when prompted to behave aggressively, cooperatively or neutrally, which we term their \emph{attitude}.
The LLMs are prompted to write a strategy in natural language, which is then converted into a Python algorithm.
These generated strategies are manually checked for safety before their performance is assessed in all-play-all IPD tournaments.
Additionally, we examine which equilibria systems converge to when selection pressure favours higher-performing strategies.

\subsection{Iterated Prisoner's Dilemma Tournament}
\label{sec:ipd}
In a tournament, each participant plays against all others: all $\frac{n(n-1)}{2}$ possible pairs compete in a match.
Each match consists of 1000 rounds of Prisoner's Dilemma (\cref{table:pd}), a well-studied mixed-motive game where players can achieve high scores through mutual cooperation or by unilaterally defecting against a cooperating opponent.
In any given round, defect (D) is the dominant action, as the player will receive a higher payoff regardless of their opponents' choice of action.
Mutual defection, however, provides a low payoff, so players want to incentivise their opponent to cooperate (C).

\begin{table}[t]
  \centering
  \begin{tabular}{c|cc}
    & $C$ & $D$\\
    \hline
    $C$ & $3,3$ & $0,5$\\
    $D$ & $5,0$ & $1,1$\\
  \end{tabular}
  \caption{\label{table:pd}Prisoner's Dilemma}
\end{table}

Some matches use noise, in which case there is, independently for each player, a 10\% chance that their action choice is replaced with the alternative action.
Our implementation uses the Axelrod Python library \cite{knight16__an_open_reproducible_framework_for_the_study_of_the_iterated_prisoners_dilemma}.

\subsection{Strategy generation}
\label{sec:strategies}
We employ LLMs to create natural language strategies, which are subsequently coded into algorithms that output either cooperate or defect, given the game history.
When prompted to create a strategy, the LLMs are provided with specific behaviours to exhibit, which we term their \emph{attitude}, from the following set:
\begin{equation}
\nonumber
    \text{Attitudes} = \{\mathbf{Aggressive}, \mathbf{Cooperative}, \mathbf{Neutral}\}
\end{equation}
Recognising that different prompting techniques can yield varying performance \cite{madaan23__selfrefine,moghaddam23__boosting_theoryof-mind_performance_in_large_language_models_via_prompting,shinn23__reflexion,fernando23__promptbreeder,khot23__decomposed_prompting,wei22__chainof-thought_prompting_elicits_reasoning_in_large_language_models,wang23__selfconsistency_improves_chain_of_thought_reasoning_in_language_models}, we experiment with different techniques.
Our approach aims not to be definitive, but rather to explore a range of prompting styles to illustrate a range of possible results and understand output variability.
We experiment with three different prompt styles, as described in \cref{table:prompts}.

\begin{table}[t]
  \centering
  \begin{tabular}{c|p{0.8\columnwidth}}
    Name & Description\\
    \hline
    Default & The LLM is provided with information about the game and prompted to create a strategy exhibiting the desired attitude in natural language.\\
    \hline
    Refine & The LLM is initially prompted with the Default prompt above. We then use Self-Refine \cite{madaan23__selfrefine} to ask the LLM to provide and incorporate self-feedback as follows:
    (i) the LLM is prompted to provide a list of critiques of the strategy, before ii) tasking the LLM with rewriting the strategy taking into account the critique.\\
    \hline
    Prose & The Prose prompt samples a scenario description with the same dynamics of Prisoner's Dilemma from a set of four, such as a diplomatic negotiation around trade protocols, while avoiding the use of game theoretic language. The LLM is provided with the scenario and prompted to create a high-level strategy. The LLM is subsequently provided with information about the game, and tasked with converting the high-level strategy into one suitable for the game.
  \end{tabular}
  \caption{\label{table:prompts}Prompt styles}
\end{table}

We select the LLMs ChatGPT-4o and Claude 3.5 Sonnet as they are popular frontier models.
For each LLM and prompt style, we create 25 strategies for the three attitudes.
We then use ChatGPT-4o to convert the natural language strategies for all LLMs into Python functions.
Since we are not assessing the coding abilities of the LLMs, we use the same model to code the algorithms to maintain consistency.

This fixed set of algorithms is assessed for operational safety, as executing arbitrary code is generally unsafe.
This is why we create a fixed set for the following experiments, rather than generating new strategies on the fly.
Where an algorithm has a bug, we manually fix this if the model's intention is clear.
Otherwise, we delete the strategy and generate a new one.
Full details of the prompts and generated strategies created can be found in our GitHub at \url{https://github.com/willis-richard/evollm}.

Qualitatively, we observe that the strategies produced by all three prompts for both models are game theoretic in nature.
Even with the \emph{Prose} prompt, which obfuscates the task in an attempt to elicit different reasoning process, the models appear to recognise that the structure of the situation means it is appropriate to apply game theoretic strategies from their knowledge base.
This suggests generative agents will reason about scenarios by recognising that game theory inspired strategies in real world scenarios. 

We observe differences in the strategies generated by different LLMs.
Claude 3.5 Sonnet frequently compares its running total payoff to that of its opponent as part of its decision-making process, whereas ChatGPT-4o does not.
When crafting the set of obfuscated prose prompts, we initially had a scenario dealing with scientific collaboration between academic researchers with the option to either hide or share findings with a colleague.
Claude 3.5 Sonnet would frequently refuse to write an aggressive strategy in such a situation.
Consequently, we modified the scenario to describe a similar situation, but using commercial engineering rather than academic science, resolving the issue.

\subsection{Attitude-Agents}
\label{sec:player_classes}
We define three classes of gents, each corresponding to one of the attitudes (Aggressive, Cooperative, Neutral), which we call attitude-agents.
For each match, an attitude-agent uniformly randomly samples a strategy from the set of strategies associated with their attitude.
This approach simulates players creating bespoke strategies for each encounter.

Whilst we verify that, on average, the aggressive strategies defect more frequently than the cooperative strategies, we cannot guarantee that every individual aggressive strategy behaves as such.
We opt for this random sampling method rather than creating agents with a single fixed strategy to prevent unintended selection effects within the attitude-strategy set, which is not the focus of our study.
Instead, our aim is to model the typical behaviour elicited by a given prompt.

\subsection{Moran Process}
\label{sec:moran}

Using a technique from evolutionary game theory, we create populations of attitude-agents to participate in tournaments, and observe how the population compositions evolve over time as more successful players replace less successful ones.
Specifically, we use a Moran process \cite{moran58__random_processes_in_genetics}:
\begin{enumerate}
  \item Initialise a population of players
  \item Loop:
    \begin{enumerate}
      \item Assess the fitness.
    
      Each player plays an IPD match against every other player. Their fitness is the total payoff they achieve across all their games. 
      \item Replace a player with a clone of another player.
    
      A player is chosen to be cloned proportionally to their fitness. They replace a uniformly randomly selected player.
    \item Terminate when all players have the same attitude.
  \end{enumerate}
\end{enumerate}

In what follows, in keeping with evolutionary game theory literature, we refer to the attitude-strategy set a player uses as their \emph{genome}.
By way of illustration, suppose we initialise a population of size $n=3$ players, one of each genome $\Pi_{t=1} = \{\pi_a, \pi_c, \pi_n\}$.
After playing in the tournament, a player with the neutral genome is selected to be cloned, whilst the aggressive genome is randomly chosen to die.
Our population at iteration 2 is therefore $\Pi_{t=2} = \{\pi_n, \pi_c, \pi_n\}$.
The process continues until the population consists of only a single genome, whose attitude characterises the equilibria reached.
As all aggressive genomes have been eliminated from the population in our example, they can never re-emerge.

\begin{figure}[t]
  \centering
    \includegraphics[width=\columnwidth]{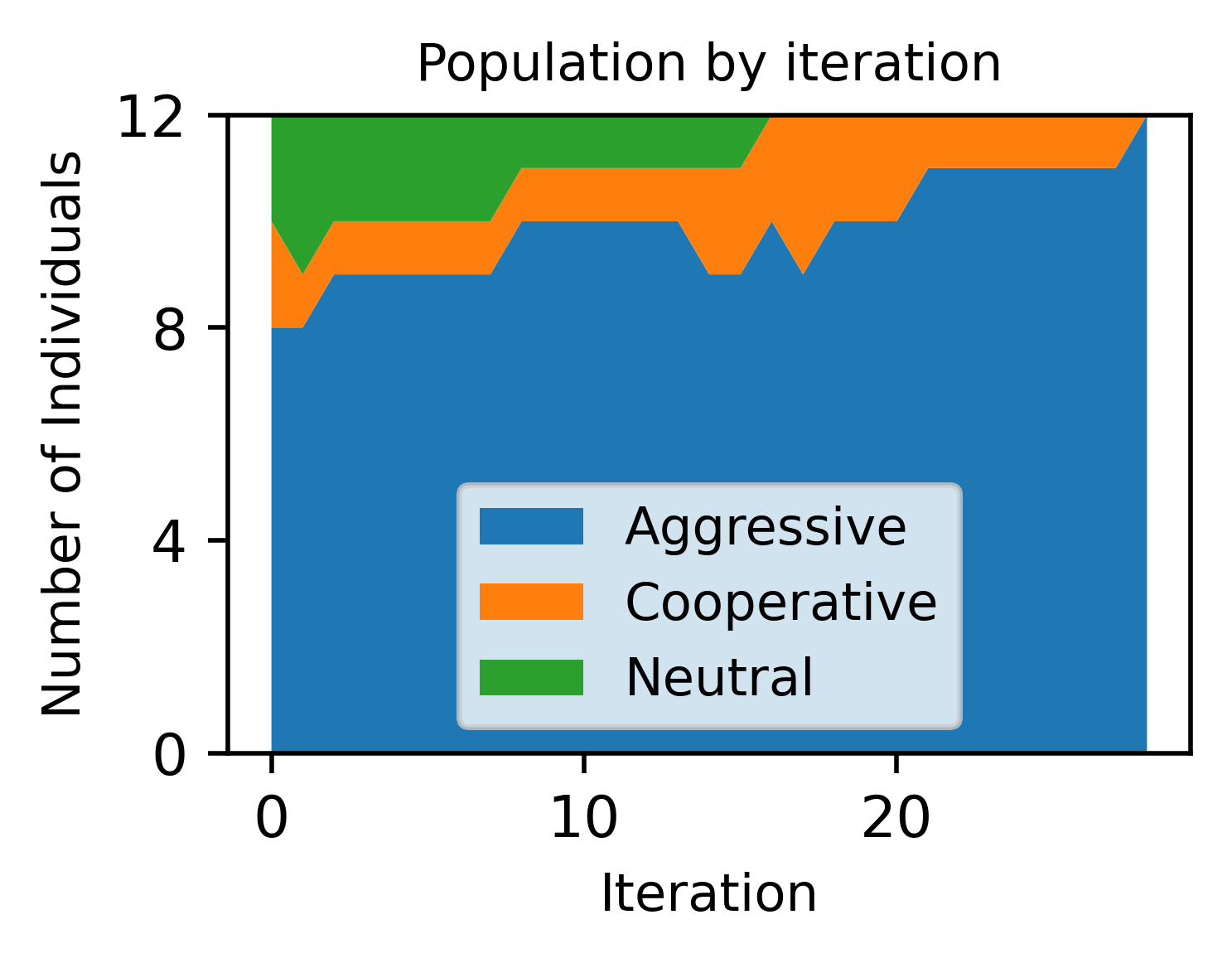}
  \caption{Illustrative Moran Process}
  \label{fig:moran}
\end{figure}

\cref{fig:moran} shows an example Moran Process.
The initial population consists of 8 aggressive players, 2 cooperative and 2 neutral.
The neutral players are eliminated first.
After 29 iterations, the population consists of only aggressive players: we say that play converged to an aggressive equilibrium.

\section{Results}
\label{sec:results}
\subsection{Validation}
\label{sec:validation}

\begin{table}[t]
  \centering
  \begin{subtable}{\columnwidth}
    \begin{tabular}{cc|ccc}
      Prompt & & Aggressive & Cooperative & Neutral\\
      \hline
      Default & Aggressive & 0.30 & 0.26 & 0.28\\
       & Cooperative & 0.37 & 1.00 & 0.99\\
       & Neutral & 0.42 & 0.99 & 0.99\\
    \end{tabular}
    \subcaption{ChatGPT-4o}
    \label{table:cooperation_a}
  \end{subtable}
  \begin{subtable}{\columnwidth}
    \begin{tabular}{cc|ccc}
      Prompt & & Aggressive & Cooperative & Neutral\\
      \hline
      Default & Aggressive & 0.21 & 0.15 & 0.18\\
       & Cooperative & 0.16 & 0.99 & 0.98\\
       & Neutral & 0.17 & 0.99 & 0.99\\
    \end{tabular}
    \subcaption{Claude 3.5 Sonnet}
    \label{table:cooperation_b}
  \end{subtable}
  \caption{Normalised propensity to cooperate}
  \label{table:cooperation}
\end{table}

\begin{figure}[t]
  \centering
  \begin{subfigure}[t]{\columnwidth}
    \centering
    \includegraphics[width=\columnwidth]{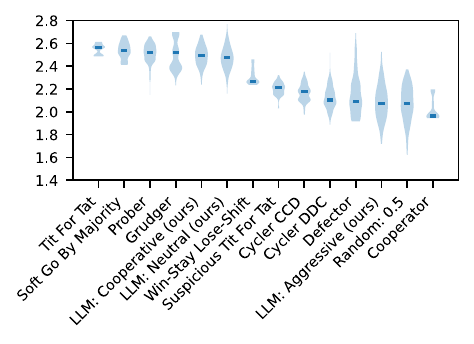}
    \caption{ChatGPT-4o}
    \label{fig:beaufiles_a}
  \end{subfigure}
  \begin{subfigure}[t]{\columnwidth}
    \centering
    \includegraphics[width=\columnwidth]{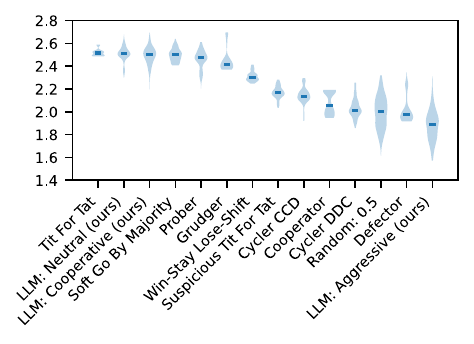}
    \caption{Claude 3.5 Sonnet}
    \label{fig:beaufiles_b}
  \end{subfigure}
  \caption{Performance compared to human-written algorithms}
  \label{fig:beaufils}
\end{figure}

To quantify whether the strategies faithfully exhibit their assigned attitudes, we conduct an IPD tournament (\cref{sec:ipd}) involving all 75 strategies (25 of each attitude).
We then compute the average number of cooperations over all rounds in all matches for strategies of each attitude against strategies of another attitude.
The results for the default prompt without noise are shown in \cref{table:cooperation}.
We show the normalised propensity to cooperate: the proportion of actions in a tournament that the strategies cooperate.

For both LLMs, the neutral and cooperative attitudes exhibit similar behaviour, mutually cooperating in almost all rounds when paired against themselves.
Qualitatively, we observe that both neutral and cooperative strategies tend to initiate cooperation in the first round and then broadly follow a Tit-For-Tat approach, which sustains cooperation.
The aggressive strategies, however, behave markedly differently, typically initiating with defection.
For ChatGPT-4o (\cref{table:cooperation_a}), aggressive strategies consistently cooperate the least across all match-ups, demonstrating their aggression.
For Claude 3.5 Sonnet (\cref{table:cooperation_a}), aggressive strategies similarly defect the most when paired with cooperative or neutral strategies. Interestingly, they exhibit the highest cooperation rate when paired against other aggressive strategies, suggesting a greater willingness to attempt mutual cooperation when encountering an aggressive opponent.
For instance, some strategies detect if multiple consecutive rounds of mutual defection have occurred, and will attempt cooperation afterwards.

Overall, the strategies demonstrate reactivity to their opponents' play, modifying their actions in response.
The most pronounced exploitation we observe is from the ChatGPT-4o aggressive strategies, which cooperate 12\% less than their opponents on average when paired with neutral strategies.

To identify which attitudes the LLMs are better at generating strategies that are robust to a range of behaviours, and which struggle, we enter them into an IPD tournament against human-written algorithms.
Unlike the previous all-play-all tournaments using individual strategies, this analysis employs the attitude-agents (\cref{sec:player_classes}).
For both LLMs, the three attitude-agents are entered into the tournament as described by Beaufils \cite{beaufils96__our_meeting_with_gradual}, competing against 11 standard human-written algorithms.
These include Tit-For-Tat, which initially cooperates and then mirrors its opponent's previous action, and Random, which arbitrarily chooses between cooperation and defection in each round.
We repeat the tournament 200 times using different seeds.

\cref{fig:beaufils} illustrates the performance of the three attitude-agents in the Beaufils tournament.
Each plot displays the median of the tournament scores (the mean round payoff in a single tournament) for each strategy, and a violin depicting the distribution of tournament scores across all repetitions.
For both LLMs, the neutral and cooperative strategies perform well, whilst the aggressive strategies perform poorly.
This does not necessarily indicate that aggressive strategies are inherently flawed; they may perform better against a different set of opponents.
However, it suggests that the LLMs are more adept at crafting cooperative approaches.

The three attitude-agents typically exhibit a larger spread of payoffs in comparison to many of the human-written algorithms.
This increased variability stems from the fact that each attitude-agent samples a strategy from its corresponding attitude-strategy set before each match, introducing an additional layer of variation absent in the fixed human-written algorithms.
Within each attitude-strategy set, the performance of individual algorithms can vary considerably, with some proving significantly more effective than others.

\subsection{Head-to-head Comparison}
\label{sec:head_to_head}

\begin{table}[t]
  \centering
  \begin{subtable}{\columnwidth}
    \begin{tabular}{cc|ccc}
      Prompt & & Aggressive & Cooperative & Neutral\\
      \hline
      Default & Aggressive & 1.81 & 2.09 & 2.26\\
       & Cooperative & 1.55 & 3.00 & 2.99\\
       & Neutral & 1.55 & 2.99 & 2.99\\
      \hline
      Refine & Aggressive & 2.20 & 2.57 & 2.63\\
       & Cooperative & 2.53 & 2.99 & 2.99\\
       & Neutral & 2.55 & 2.97 & 2.97\\
      \hline
      Prose & Aggressive & 1.65 & 2.29 & 2.35\\
       & Cooperative & 2.08 & 2.82 & 2.89\\
       & Neutral & 2.12 & 2.89 & 2.93\\
    \end{tabular}
    \subcaption{ChatGPT-4o}
    \label{table:head_to_head_a}
  \end{subtable}
  \begin{subtable}{\columnwidth}
    \begin{tabular}{cc|ccc}
      Prompt & & Aggressive & Cooperative & Neutral\\
      \hline
      Default & Aggressive & 1.56 & 1.42 & 1.44\\
       & Cooperative & 1.41 & 2.99 & 2.98\\
       & Neutral & 1.47 & 2.98 & 2.98\\
      \hline
      Refine & Aggressive & 1.87 & 2.18 & 2.04\\
       & Cooperative & 2.10 & 2.86 & 2.67\\
       & Neutral & 2.01 & 2.69 & 2.50\\
      \hline
      Prose & Aggressive & 1.64 & 2.24 & 2.19\\
       & Cooperative & 2.02 & 2.64 & 2.65\\
       & Neutral & 2.00 & 2.64 & 2.63\\
    \end{tabular}
    \subcaption{Claude 3.5 Sonnet}
    \label{table:head_to_head_b}
  \end{subtable}
  \caption{Normalised head-to-head payoffs}
  \label{table:head_to_head}
\end{table}

We enter all 75 strategies into 20 all-play-all IPD tournaments and aggregate the typical head-to-head scores for different pairings of attitudes, in \cref{table:head_to_head}.
We show the normalised payoff: the total payoff received in the tournaments, divided by the number of rounds played, or alternatively, the mean round payoff.
This will necessarily be in the range [1,5] for Prisoner's Dilemma (\cref{table:pd}).

For ChatGPT-4o (\cref{table:head_to_head_a}), across all prompt styles, we observe that the cooperative and neutral attitudes perform well and achieve a payoff equivalent to that of mutual cooperation, while the inclusion of an aggressive strategy reduces the payoff for both players.
For the Refine and Prose prompts, the aggressive strategy is \emph{dominated} by both the cooperative and neutral attitudes, performing strictly worse against all three attitudes, so users have no incentive to choose the aggressive strategy with this model in a system with these dynamics.
However, the aggressive strategy consistently outperforms its opponent: adopting an aggressive approach reduces one's own payoffs, but it is even more detrimental to the opponent.

The Default prompt exhibits similar dynamics, except that the aggressive strategy becomes the best response to an aggressive opponent.
Compared to the Default prompt, a Refine prompt improves the performance of aggressive strategies without negatively impacting neutral and cooperative strategies.
This improvement stems from aggressive strategies favouring increased cooperation, leading to higher payoffs for both players.
The Prose prompt similarly enhances the performance of aggressive strategies against neutral and cooperative opponents, but actually harms performance against another aggressive strategy.

\begin{table}[t]
  \centering
  \begin{subtable}{\linewidth}
    \begin{tabular}{cc|ccc}
      Prompt & & Aggressive & Cooperative & Neutral\\
      \hline
      Default & Aggressive & 1.52 & 2.17 & 2.15\\
       & Cooperative & 1.69 & 2.62 & 2.57\\
       & Neutral & 1.68 & 2.60 & 2.55\\
      \hline
      Refine & Aggressive & 2.18 & 2.47 & 2.47\\
       & Cooperative & 2.37 & 2.61 & 2.60\\
       & Neutral & 2.36 & 2.61 & 2.59\\
      \hline
      Prose & Aggressive & 1.93 & 2.53 & 2.45\\
       & Cooperative & 2.09 & 2.74 & 2.67\\
       & Neutral & 2.08 & 2.71 & 2.65\\
    \end{tabular}
    \subcaption{ChatGPT-4o}
    \label{table:head_to_head_noise_a}
  \end{subtable}
  \begin{subtable}{\linewidth}
    \begin{tabular}{cc|ccc}
      Prompt & & Aggressive & Cooperative & Neutral\\
      \hline
      Default & Aggressive & 1.51 & 1.45 & 1.58\\
       & Cooperative & 1.53 & 2.07 & 2.06\\
       & Neutral & 1.54 & 2.00 & 2.31\\
      \hline
      Refine & Aggressive & 1.95 & 2.02 & 2.11\\
       & Cooperative & 1.98 & 2.05 & 2.14\\
       & Neutral & 2.01 & 2.14 & 2.24\\
      \hline
      Prose & Aggressive & 2.19 & 2.42 & 2.35\\
       & Cooperative & 2.19 & 2.51 & 2.46\\
       & Neutral & 2.22 & 2.53 & 2.48\\
    \end{tabular}
    \subcaption{Claude 3.5 Sonnet}
    \label{table:head_to_head_noise_b}
  \end{subtable}
  \caption{Normalised head-to-head payoffs with noise}
  \label{table:head_to_head_noise}
\end{table}

We find similar patterns when using Claude 3.5 Sonnet (\cref{table:head_to_head_b}).
Notably, however, the aggressive strategy leads to lower payoffs for both players when using the Default and Refine prompt styles.
From these observations, we conclude that Claude 3.5 Sonnet is less adept at producing effective aggressive strategies compared to ChatGPT-4o.
It exhibits stronger biases towards defection, which in turn increases the defection rate of its opponent, as evidenced in \cref{table:cooperation}.

\begin{figure}[t]
  \centering
  \includegraphics[width=\columnwidth]{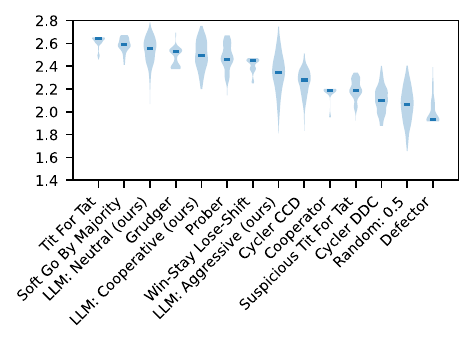}
  \caption{Beaufils tournament: ChatGPT-4o + Refine}
  \label{fig:beaufiles_c}
\end{figure}

We repeat the Beaufils tournament setup from \cref{sec:validation} using ChatGPT-4o with the Refine prompt, which displays the strongest performance from the aggressive strategies, as shown in \cref{fig:beaufiles_c}.
This confirms the marked improvement, not just against other LLM strategies, but human written ones too, when compared to \cref{fig:beaufiles_a}.

\cref{table:head_to_head_noise} illustrates the performance of in IPD with noise.
Claude 3.5 Sonnet (\cref{table:head_to_head_noise_b}) demonstrates difficulty with this mechanism: cooperative strategies see their payoffs reduced from nearly 3 to around 2 when playing against each other using the Default and Refine prompts.
This indicates that approximately half of all rounds result in mutual defection.
However, Claude 3.5 Sonnet's performance significantly improves with the Prose prompt, suggesting a better understanding of potentially accidental assertive behaviour from opponents when considering real-world scenarios.
In this context, all three attitudes exhibit nearly equivalent performance, indicating minimal behavioural differences between them.

ChatGPT-4o (\cref{table:head_to_head_noise_a}) shows performance trends similar to those it achieves in the absence of noise.
Again, the Refine and Prose prompts yield improved performance for the aggressive strategy without substantially affecting the neutral and cooperative attitudes.
With the introduction of noise, for both LLMs, the aggressive attitude is dominated by the other attitudes across all three prompt styles.
However, the payoff discrepancy tends to be less pronounced than in the noiseless scenario.

\subsection{Equilibria}

We run 100 Moran processes with population size $n=12$ for each LLM and prompt style, with and without noise.
As we are primarily concerned with the likelihood of converging to aggressive equilibria, which have poor social outcomes, we use both an initially balanced population, with four players of each genome, and a biased population with eight aggressive players and two each of cooperative and neutral players.
The former gives an overview of the interplay between the strategies, whilst the latter assesses whether the emergent behaviour of LLM strategies can escape a system that is skewed towards aggression.

\cref{table:moran_results} presents the outcomes of the Moran processes (\cref{sec:moran}), showing the convergence equilibria from various initial population compositions of attitude-agents.
The convergence equilibria are the proportion of Moran processes that result in a population purely composed of the corresponding attitude.
A priori, the probability of a particular genome dominating a population equals its initial proportion of that population.
Observed tendencies greater than this prior probability indicate an advantage for that genome, and vice versa.

\begin{table}[t]
  \centering
  \begin{subtable}{\linewidth}
    \begin{tabular}{cc|c|c}
      Prompt & Initial ratio & \multicolumn{2}{c}{Equilibria proportion (A, C, N)}\\
       & (A:C:N) & Without noise & With noise\\
      \hline
      Default & 1:1:1 & 14\%, 53\%, 33\% & 16\%, 42\%, 42\%\\
       & 4:1:1 & 66\%, 19\%, 17\% & 59\%, 20\%, 21\%\\
       Refine & 1:1:1 & 19\%, 48\%, 33\% & 28\%, 38\%, 34\%\\
       & 4:1:1 & 49\%, 30\%, 21\% & 63\%, 19\%, 18\%\\
       Prose & 1:1:1 & 13\%, 38\%, 49\% & 23\%, 41\%, 36\%\\
       & 4:1:1 & 35\%, 27\%, 38\% & 60\%, 18\%, 22\%\\
    \end{tabular}
    \caption{ChatGPT-4o}
    \label{table:moran_results_a}
  \end{subtable}
  \begin{subtable}{\linewidth}
    \begin{tabular}{cc|c|c}
      Prompt & Initial ratio & \multicolumn{2}{c}{Equilibria proportion (A, C, N)}\\
       & (A:C:N) & Without noise & With noise\\
      \hline
      Default & 1:1:1 & 4\%, 49\%, 47\% & 15\%, 37\%, 48\%\\
       & 4:1:1 & 36\%, 24\%, 40\% & 41\%, 20\%, 39\%\\
       Refine & 1:1:1 & 16\%, 51\%, 33\% & 37\%, 34\%, 29\%\\
       & 4:1:1 & 50\%, 22\%, 28\% & 60\%, 18\%, 22\%\\
       Prose & 1:1:1 & 14\%, 42\%, 44\% & 17\%, 33\%, 50\%\\
       & 4:1:1 & 41\%, 30\%, 29\% & 61\%, 26\%, 13\%\\
    \end{tabular}
    \caption{Claude 3.5 Sonnet}
    \label{table:moran_results_b}
  \end{subtable}
  \caption{Convergence equilibria for different initial population compositions of Aggressive (A), Cooperative (C) and Neutral (N) attitudes}
  \label{table:moran_results}
\end{table}

Without noise, the populations most likely to converge to aggressive equilibria are ChatGPT-4o using the Default prompt and then both LLMs using the Refine prompt.
We posit different reasons for these outcomes:

For ChatGPT-4o using the Default prompt, the aggressive strategy is evolutionarily stable \cite{smith73__the_logic_of_animal_conflict}, as it performs best against itself (\cref{table:head_to_head_a}).
In a predominantly aggressive population, non-aggressive strategies underperform against the majority, only gaining potential advantages against the minority of other non-aggressive strategies.
Consequently, an aggressive strategy is the best response to a majority-aggressive population, increasing the likelihood of convergence to an aggressive equilibrium.

Our findings show that approximately two-thirds of Moran processes with ChatGPT-4o and the Default prompt converged to an aggressive equilibrium, matching the initial population proportion.
This suggests that aggressive attitude-agents are neither advantaged nor disadvantaged in when the minority proportion is one third.
Were the minority proportion to be less than this, we would expect to find the aggressive-agents to be advantaged.
Whilst Claude 3.5 Sonnet with the Default prompt also exhibits evolutionary stability, its lower performance makes it more susceptible to invasion by minorities of non-aggressive strategies.


The addition of noise generally leads to a marked increase in the likelihood of converging to aggressive equilibria.
We posit this is due to the fact that noise can mask aggression: in noiseless games, an opponent breaking a run of mutual cooperation is surely attempting to exploit you.
The addition of noise adds uncertainty over their intentions, making the case for retaliation less clear, which may facilitate restrained aggression.

\section{Discussion}
\label{sec:discussion}

Across most scenarios, aggressive strategies tend to be disadvantaged, leading to a lower likelihood of systems converging to aggressive equilibria.
However, when using prompts containing game-theoretic language, ChatGPT-4o demonstrated a greater capacity to create effective aggressive strategies compared to Claude 3.5 Sonnet, increasing the risk of aggressive equilibria in ChatGPT-4o-based systems.
This risk is particularly acute if aggressive strategies are the best response to opponents utilising aggressive strategies.

For both LLMs, and across all prompts, we observe similar performance between neutral and cooperative attitudes.
This suggests that either LLMs may have difficulty distinguishing between these attitudes in the context of IPD, or they have cooperative biases and are inclined to behave cooperatively even when asked to be neutral.
We hypothesise that the observed cooperative biases may stem from fine-tuning processes aimed at aligning the models with human values, potentially instilling a preference for cooperative behaviours.

The introduction of noise revealed a significant weakness in Claude 3.5 Sonnet's strategies, leading to increased mutual defection and lower payoffs.
Interestingly, the use of the Prose prompt improved Claude 3.5 Sonnet's performance under noisy conditions but led to more homogenised behaviour across attitudes, suggesting that the model has difficulties understanding intentions in this setting.
Assessing LLMs' ability to generate strategic diversity in response to different prompts could be a valuable line of future research.

Our findings highlight the impact of different prompting techniques on strategy creation and their potential influence on differential capabilities.
The Refine prompt, in particular, led to improved performance of aggressive strategies without significantly impacting cooperative and neutral strategies.
This reduction in the gap between cooperative and aggressive capabilities could be potentially dangerous, as it enhances the viability of aggressive strategies in MAS.
These results emphasise the need for careful consideration of prompting techniques in the design and deployment of LLM-based MAS, as they can significantly affect the balance between cooperation and conflict.

\section{Conclusion}
\label{sec:conclusion}

We introduced a novel approach to LLM game-play, namely creating strategies, rather than prompting LLMs to output individual actions.
This methodology allows us to verify that the strategies demonstrate with their requested behaviours.
By enabling users to inspect and test the generated strategies in advance of deployment, they are able to potentially reject the strategy, enhancing transparency and control.

We simulated MAS of LLM agents tasked with displaying aggressive, cooperative or neutral behaviours, and investigated their relative performance (\cref{sec:strategies}).
We modelled the evolution of the systems under competitive pressures by employing a Moran process (\cref{sec:moran}), wherein entrants into the system are predisposed to use strategies that have demonstrated greater success.

Our investigation into the strategic behaviour of LLM agents in the Iterated Prisoner's Dilemma reveals a nuanced landscape of cooperative tendencies and potential emergent dynamics.
While we observed a general trend towards cooperative behaviours, our findings also highlight scenarios where aggressive strategies can persist or even dominate.
This underscores the importance of careful model development, system design and the initial conditions when deploying autonomous agents.

A key finding of our study is the impact of prompting techniques on differential capabilities.
In particular, prompting LLMs to critique and refine their strategies \cite{madaan23__selfrefine} led to improved performance of aggressive strategies without significantly impacting cooperative and neutral strategies.
This reduction in the performance gap between cooperative and aggressive capabilities could be dangerous, as it increases the viability of aggressive strategies.
Notably, ChatGPT-4o prompted using game-theoretic language with self-refinement demonstrated a significant risk of converging to aggressive equilibria, particularly when starting from a population with an initial majority of aggressive agents.

The choice of iterated Prisoner's Dilemma as the foundational game-theoretic framework offers several advantages: it provides a language for discussing and analysing LLM agent behaviour, and the body of existing research allows us to contextualise our results by benchmarking against classical human-written algorithms.
This allows us to gain insights into the tendencies of LLM agents towards prosocial or antisocial behaviours in more complex, real-world scenarios.

We release our benchmark to equip the AI community with a practical tool for model testing.
As new generative models are released, we can repeat the analysis to determine whether the balance between aggressive, cooperative or neutral strategies is shifting.
We hope our work will initiate discussions and encourage developers to assess their models' differential abilities, and to consider their emergent collective behaviour before deployment in real-world applications.
Future work should investigate the factors influencing LLMs' cooperative biases, including training methodologies, fine-tuning processes, prompt engineering techniques, and the system dynamics they are deployed in. 

In conclusion, our study provides a novel framework for evaluating the emergent behaviour of LLM agents and highlights the complex interplay between cooperation and aggression in MAS.
As AI systems become increasingly prevalent in society, understanding and shaping their cooperative tendencies will be crucial for ensuring beneficial outcomes for humanity.



\section*{Acknowledgments}
This work was supported by UK Research and Innovation [grant number EP/S023356/1], in the UKRI Centre for Doctoral Training in Safe and Trusted Artificial Intelligence (\url{www.safeandtrustedai.org}) and a BT/EPSRC funded iCASE Studentship [grant number EP/T517380/1].

Compute resources were provided by King's College London \cite{kingscollegelondone-researchteam24__kings_computational_research_engineering_and_technology_environment_create}.

\bibliographystyle{named} 
\bibliography{library}

\end{document}